# End-of-Life of Software
# How is it Defined and Managed?

**White paper**

April 2022

Written by Dr Zena Assaad and Dr Mina Henein

# Abstract

The rapid development of new software and algorithms, fueled by the immense amount of data available, has made the shelf life of software products a lot shorter. With a rough estimate of more than 40,000 new software projects developed every day[1], it is becoming quicker and cheaper to abandon old software and acquire new software that meets rapidly changing needs and demands. What happens to software that is abandoned and what consequences may arise from 'throwaway' culture (Cooper, 2005) are still open questions.

This paper will explore the systems engineering concept of end-of-life for software, it will highlight the gaps in existing software engineering practices, it will bring forward examples of software that has been abandoned in an attempt to decommission and it will explore the repercussions of abandoned software artefacts. A proposed way forward for addressing the identified research gaps is also detailed.

# Introduction

Large scale physical artefacts that service many people, such as bridges, road infrastructure, oil and gas reserves, etc., are predominantly developed against systems engineering processes. This curated process guides the development of a system or product against stated needs, requirements, budget and schedule (Haskins et al., 2007). A product or system is developed through a life cycle process, starting from inception, through to planning, progressing to analysis, requirements and design, incorporating development and testing before moving into implementation or deployment and following with operations and maintenance (Day, 1981).

Systems engineering practices have been refined over many decades. The phases across a systems lifecycle are managed through tried and tested processes that have demonstrated value and success. The foundation of systems engineering practices have been translated to software development, particularly for the early stages of the systems lifecycle from the planning to development stages. What is not yet represented in software lifecycles is the final stages of a systems lifecycle.

For physical artefacts, the final stages of their life cycle involve identifying that the product has reached its end-of-life (EOL) and upon that, decommissioning or disposing of the product. The purpose of decommissioning a product is to protect it and its associated systems, to reduce ongoing costs, and to reduce potential hazards and risks the product might impose on other connected systems (Shaw, 1994).

---

[1] *Calculated as the total number of software projects on GitHub – the world's biggest software projects repository– divided by the total number of days since its launch in 2008.*

The concept of decommissioning has evolved over the last few decades. When and how a product or system is decommissioned varies widely between disciplines (Reisenweaver, 2002). However, a common thread among this broad set of disciplines is the importance of decommissioning as this practice supports sustainability of the broader system in which a product or system operates (Chatzis, 2016).

The digital age, often referred to as the information age, is characterised by a widespread shift towards digital technology. This shift has fostered increased development of more intangible artefacts and services which are increasingly being infused in our everyday lives. Intangible products and systems hold the benefit of not being vulnerable to the stressors imposed on physical artefacts, such as corrosion, degradation of materials, etc. Despite this advantage, software will still have a finite life span. Determining when a software product has reached EOL is a complex task. Currently, there is sparse and limited literature highlighting practices for the determining EOL of software. There is also a notable gap in the literature around explored or proposed practices for decommissioning software.

The concept of decommissioning has evolved over time, and with the changing industrial landscape that has come with the digital age, this concept will continue to evolve for different industries. This paper will highlight the importance of determining the EOL of software and it will explore the effects of a lack of adequate decommissioning processes for software products. This article will also propose a method for determining EOL of software and subsequently decommissioning software.

# What is End-of-Life?

EOL refers to the point at which a product or system no longer meets the original design needs or expectations (Lee et al., 2010). The term EOL can be misleading as a product that has reached this point may still be functional or may still maintain some value, despite no longer meeting needs or expectations.

There exists standard practices that have been developed and refined over the years for determining when a product has reached EOL (Herrmann et al., 2008). There are usually clear signals or attributes that indicate a product is no longer fit for purpose. Safety is one prominent attribute for determining EOL for physical artefacts (Ma et al., 2017). How safe a product is to use, particularly from the perspective of physical safety to humans, will change as a product ages. As materials age, they deteriorate and are prone to symptoms such as erosion and a reduction in load bearing capacity (Šomodíková et al., 2016). Consequently, the risk landscape will evolve as the product or system matures.

Environmental impact is another measure of EOL of a product or system. The impact that products or systems have on the environment is becoming a prominent consideration for manufacturers and consumers at large. The demand for more environmentally conscious products coupled with emerging international environmental laws (Guruswamy & Leach, 2017), has resulted in the development of products and systems with a smaller carbon footprint overall. The increased demand for more environmentally conscious products has led to discontinued manufacturing of older models. For example, motor vehicles being designed and manufactured today, are far more fuel efficient than those manufactured even a decade prior (Ross, 2010).

Another prominent attribute used to determine EOL is cost. As products age, they can become too costly to maintain. Products and systems rarely operate in isolation of other products, infrastructure or systems. For example, a bridge does not operate in isolation of motor vehicles, road infrastructure or traffic systems. The cost of one entity, within the ecosystem in which it operates, will ultimately have flow on effects to the other entities within its network. Investing in a newer version or model is often more cost effective in the long term (Rothenberg, 2007).

Measures for determining EOL of a product or system extend beyond the three examples provided and broaden for different industries and applications (Haskins et al., 2007). For non-physical artefacts, such as software, the measures or signals for determining EOL are less clear and straightforward.

# End-of-Life of Software

Software is malleable and not subject to physical degradation over time. It is a dynamic artefact that evolves with each update to meet changing user needs and preferences. The amorphous nature of software makes considerations around its lifespan more complex to unpack. Questions around measures of deterioration over time and measures of safety are more challenging to address.

Once it has been determined that a product or system has reached its EOL, it then needs to be decommissioned. Decommissioning can be broadly defined as the retirement of a product or system through dismantling, disposing or repurposing of products and materials (Haskins et al., 2007). There does not exist a singular overarching definition for decommissioning; however, there exist definitions within different industries. For example, the International Atomic Energy Agency defines decommissioning as the administrative and technical actions taken to allow the removal of some or all of the regulatory controls from a nuclear facility (Reisenweaver, 2002). The approach to and process of decommissioning will differ for different industries due to the proliferation of applications of products or systems across different sectors.

For some industries, guidelines and regulations have been established to regulate the decommissioning process (NOPSEMA, 2021), for example ISO 26262-7 which looks at functional safety for decommissioning of electrical and/or electronic systems that are installed in road vehicles (ISO, 2018). More broadly, there are no existing guidelines or regulations for decommissioning that are industry agnostic. That is, there are no unified guidelines and regulations around decommissioning that would be applicable to every industry.

To date, a majority of software products are not decommissioned, based on the broad definition of the term. For software, there is a history of ending service of the product, such as updates and supporting resources, in an attempt to retire the artefact. An example of this is the web browser "Netscape Communicator" which was released to the public in the late 1990's. The web browser was made available through an open source licence and would later be owned by the Mozilla foundation. In 2006, development of this web browser was discontinued in an effort to retire the product. As the product was made available to the public through an open source licence, the product remained in operation through a community maintained version called SeaMonkey (Jeffrey & Franco, 2020).

Netscape communicator is an example of how the notion of EOL can be misleading. While the owners of the artefact determined the product had reached its EOL, or a point of end-of-service, the product itself still had use and value among the public.

Netscape Communicator is one of many examples of platforms that go unsupported as a means of retiring the product, with this approach earning the term "abandonware" (Khong, 2006). Ceasing service of a product does not mean that the product is no longer being used. In many cases, emulators are used to run old software on modern computers, thereby extending the usability of a product. However, without the proper support, software security is compromised (McGraw, 2004).

The prevalence of abandonware has created issues in the legal circle around copyright law as continual use and development of a product by the public impeaches on laws around copying and distributing work (Copyright, 1988).

In addition to concerns around breaching copyright law, there are implications around the social impact of products that continue to be used following abandonment by their creators. A salient example of the influence the public has over publicly released products is the Microsoft chatbot Tay. In 2016 Microsoft released a chatbot, Tay, which was designed to converse with people on Twitter. The bot was intended to learn from conversations and mimic language patterns. Some users began conversing with the bot using inflammatory and racist language. Within 16 hours Microsoft suspended Tay's twitter account following more than 96,000 tweets, many of which included inflammatory statements (Hunt, 2016).

When a product is abandoned, or service for a product is discontinued, it is usually for a particular reason. The product may be too costly to maintain, newer technology supersedes the capabilities of the product, the product may not function as intended, etc. On the surface, the continuation of maintenance and use of a product by the public may seem harmless. However, as we saw with Tay, the public can have a significant influence on the outcomes and subsequent impact of software. Publicly maintained software can also be subject to issues around liability and accountability. Liability refers to a legal responsibility, this is where concerns around copyright law would emerge, and accountability refers to social responsibility, which is more subjective and nuanced. As an example, the algorithms underpinning search engines, such as Google, have underlying biases in their design (Noble, 2018). If a search engine were to be maintained by a public community and the underlying biases were magnified, who would be held accountable? Noting, it would be difficult to hold anyone legally liable as existing laws do not encompass algorithmic biases. This lends itself to two questions: who decides a software product has reached its EOL, its creators or its users? And, what responsibility do creators of software products have on how these artefacts are decommissioned once they have reached their EOL?

# Why Do We Need to Decommission Software?

Software is an amorphous artefact that can be improved upon with time through periodic updates. While there are examples of software products being retired for various reasons, there are also examples of software products that have demonstrated a considerably long lifespan. The extended lifespan of software systems has resulted in ramifications that impact its usability and accessibility over time.

The financial industry is dominated by one particular coding language, Common Business Oriented Language or COBOL. COBOL was developed in 1959 and remains in use in financial systems today, almost six decades from its first release. As a product rarely operates in isolation of other products, infrastructure or systems, the continued use of COBOL over six decades has resulted in some challenges and impacts to broader ecosystems. One of the main challenges is around the workforce required to maintain the language. COBOL is not a common coding language outside of the finance industry, resulting in a narrow emerging workforce that is expected to decline rapidly over the next decade (Fleishman, 2018; McKay, 2020). COBOL is a pertinent example of software not being discontinued due to dependence on associated third party software.

Despite being amorphous, software will still have a finite lifespan, and therefore, it is important to establish a means of identifying when a software product has reached its EOL. Currently, there are a number of factors that impact the viability of continuing to maintain and use developed software. Some of these factors include technical debt, viability of commercial off-the-shelf solutions, dependence on multiple programming languages and associated third-party libraries.

Technical debt describes the consequences of software development actions that intentionally or unintentionally prioritise client value and/or project constraints such as delivery deadlines, over more technical implementation and design considerations (Holvitie et al., 2018). In other words, it is the difference between what was promised and what was actually delivered or the price you need to pay tomorrow because you took a shortcut in order to deliver the software today.

While there are many contributing factors to technical debt (Product Plan, 2021), a lack of documentation, known as documentation debt, is one of the main contributing factors. An ill-documented project might result in the need to rewrite the whole project in the event the development team changes, resulting in added cost and time.

Another factor that impacts the viability of continuing to maintain and use a software product is the dependence of the software on multiple programming languages and associated third-party libraries and software. An individual or a company might choose to replace a software product because the associated systems (software or hardware) have been discontinued, are no longer maintained or have been replaced with the latest state-of-the-art products to which the software is no longer adequate.

Once a product or service has been determined to have reached its EOL, however that is defined, it then needs to be decommissioned or disposed of. For traditional physical artefacts, this process often involves breaking down the artefact and disposing of its parts following known regulations and guidelines. For software, an intangible artefact composed of machine code, the concept of decommissioning or disposal needs to be re-evaluated.

The purpose of decommissioning a software product is to protect it and its associated systems, to reduce ongoing costs, and to reduce potential hazards and risks the product might impose on other connected systems (Fortune & Paterson, 2018). These attributes are what distinguish decommissioning from abandonment. With the high dependency of newly developed software on large data sets, each piece of code encodes information about people, places, tools, and scenarios that might represent risks if not disposed of properly. Questions around risks and privacy are now, more than ever, relevant to software decommissioning. Decommissioning of a software product not only minimises associated risks, but it also ensures continuity and sustainability of the broader system.

# Next Steps

This project aims to identify attributes to determine the EOL of software products as a step towards establishing standardised ways and procedures to decommission or dispose of such products.

It can be difficult to ascertain attributes indicating that software has reached its EOL. This project intends to explore measures for determining EOL of software. Case studies which detail examples of software products that have been discontinued or deemed no longer fit for purpose will be analysed, with the aim of determining when and why the use of such products were deemed no longer fit for purpose.

In addition to identifying attributes to determine the EOL of software products, this project also aims to develop methodologies that can be used to assess if a software product has reached its EOL. While the format of such methodologies is unknown at this stage, developing a tool that would be used by software users and developers to assess if a certain product has reached its EOL while accounting for its user input as a feedback mechanism is envisaged. This tool is intended to be a community based tool, made openly accessible alongside supporting documentation detailing how to use the tool.

Case study analysis will be coupled with workshops with subject matter experts. The workshops will include a diversity of experts including software developers, software users, systems engineers, etc. These workshops will build on the case study analysis through collaborative investigation of the outputs of those analysis, with the purpose of validating those outcomes.

Following on from this, this project will look at defining decommissioning for software and outlining processes for decommissioning software. Decommissioning is defined differently across the broad set of industries that employ systems engineering practices. It is not yet clearly defined for software. Before undertaking the task of determining how software can be decommissioned, providing clarity on what it means to decommission software is needed.

This project is a look into how we can identify that a product should no longer be used. It will explore what signals or attributes we can use to determine a product has reached its EOL and what steps should be taken to ensure that product is decommissioned or disposed of efficiently.

# Conclusion

As the capability and potential applications of software continue to broaden, the demand for these products is growing exponentially. Software products are quickly becoming the infrastructure that underpins many industries. The advancements in software are surpassing the development of guidance and standards around best practice processes. While there are examples of systems engineering practices being translated to software development, these examples reflect the early stages of the lifecycle process. There exists a gap in the research around practices and processes for determining EOL and decommissioning software. This gap has led to complacency around when and how software is decommissioned. As software is becoming increasingly easier to replace, the measures for determining a product is no longer fit for purpose are unclear and inconsistent. Furthermore, because of the lack of guidance around how to decommission software, the phenomena of abandonware has emerged. Research is needed to explore what attributes can be used to measure EOL of software and what practices can be pursued to ensure software is decommissioned effectively and responsibly.

It? *How-To Geek.* https://www.howtogeek.com/667596/what-is-cobol-and-why-do-so-many-institutions-rely-on-it/

19. Noble, S. (2018). *Algorithms of oppression: How search engines reinforce racism*. New York University Press.

20. NOPSEMA. (2021). *Decommissioning* (National Offshore Petroleum Safety and Environmental Management Authority). https://www.nopsema.gov.au/offshore-industry/decommissioning

21. Product Plan. (2021). Technical Debt. *Product Plan*. https://www.productplan.com/glossary/technical-debt/

22. Reisenweaver, D. W. (2002). The IAEA'S Decommissioning Concept. *Safe Decommissioning for Nuclear Activities*, 123–130. https://www-pub.iaea.org/MTCD/publications/PDF/Pub1154_web.pdf#page=130

23. Ross, M. (2010). Fuel efficiency and the physics of automobiles. *Contemporary Physics*, *38*(6), 381–394. https://doi.org/10.1080/001075197182199

24. Rothenberg, S. (2007). Sustainability Through Servicizing. *MIT Sloan Management Review*, *48*(2), 83–91.

25. Shaw, K. (1994). *Decommissioning and Abandonment: The Safety and Environmental Issues.* SPE-27235-MS. https://doi.org/10.2118/27235-MS

26. Šomodíková, M., Lehký, D., Doležel, J., & Novák, D. (2016). Modeling of degradation processes in concrete: Probabilistic lifetime and load-bearing capacity assessment of existing reinforced concrete bridges. *Engineering Structures*, *119*, 49–60. https://doi.org/10.1016/j.engstruct.2016.03.065
Z. Assaad, M. Henein                                                                                               Page 13